# On Close Relationship between Classical Time-Dependent Harmonic Oscillator and Non-Relativistic Quantum Mechanics in One Dimension

Alexander Davydov

AlgoTerra LLC, 249 Rollins Avenue, Suite 202, Rockville, MD 20852, U.S.A.
E-mail: alex.davydov@algoterra.com

**Abstract.** In this paper I present a mapping between representation of some quantum phenomena in one dimension and behavior of a classical time-dependent harmonic oscillator. For the first time, it is demonstrated that quantum tunneling can be *described* in terms of classical physics without invoking violations of the energy conservation law at any time instance. A formula is presented that generates a wide class of potential barrier shapes with the desirable reflection (transmission) coefficient and transmission phase shift along with the corresponding *exact* solutions of the time-independent Schrödinger's equation. These results, with support from numerical simulations, strongly suggest that two uncoupled classical harmonic oscillators, which initially have a 90° relative phase shift and then are simultaneously disturbed by the same *parametric perturbation of a finite duration*, manifest behavior which can be mapped to that of a single quantum particle, with classical 'range relations' analogous to the uncertainty relations of quantum physics.

**PACS numbers:** 03.65.Ge, 03.65.Sq, 03.65.Xp

**Keywords:** Quantum Mechanics • Time-Dependent Harmonic Oscillator • Uncertainty Relations • Tunneling

## 1 Introduction

The fundamental concepts and mathematical formalism of quantum mechanics are significantly different from those of its classical counterpart resulting in non-intuitive behavior of quantum objects without models that are as easy to visualize as those in the classical world. According to the general belief, quantum mechanics is the fundamental theory and the laws of motion of macroscopic bodies emerge from it in the appropriate limit, although consensus is yet lacking on details of how exactly this happens. Considerable amount of quantum-classical analogies have been identified [1]. However, the conventional wisdom among physicists is that these analogies are mere mathematical curiosities and the formalism of classical mechanics is incapable of describing many purely quantum phenomena such as tunneling, wave-particle duality, or interference. In this paper, I challenge this point of view by showing that, at least in the case of one dimension, the relationship between quantum and classical physics is much deeper than is currently acknowledged. Specifically, I demonstrate that quantum tunneling can be described in terms of classical physics via mapping that links a complex-valued wavefunction – solution of the *time-independent* Schrödinger's equation (TISE) – with classical behavior of a pair of uncoupled *time-dependent* harmonic oscillators (TDHOs). I also present the numerical evidence supporting the conjecture that there exist classical analogs of the time-energy and position-momentum uncertainty relations for a macroscopic TDHO parametrically perturbed during finite interval.

The paper is organized as follows. In section 2, some known (but not widely recognized) facts related to one-dimensional TDHO and its equation of motion are briefly introduced. These particulars are essential for comprehending the rest of the paper. Section 3 presents a novel method of finding a pair (TDHO equation, its exact solution in analytic form) which can be used for generating many previously unknown such pairs. The main result – close relationship between non-relativistic quantum mechanics and physics of a classical TDHO in one dimension – is deduced in section 4. Section 5 provides numerical evidence in support of the conjecture about existence of the *range relations* for a macroscopic TDHO analogous to the time-energy and position-momentum uncertainty relations of quantum mechanics. This analysis is based on exact solutions of the TDHO equation. In section 6, the isomorphism between quantum tunneling and entirely classical behavior of a TDHO is discussed, and the formula is deduced that generates a wide class of one-dimensional barrier shapes in analytic form with the desirable reflection (transmission) coefficient and transmission phase shift for impinging flow of quantum particles with definite energy. As an illustration of the capabilities of this technique, three *exact* solutions of the time-independent Schrödinger's equation are presented, all of them previously unknown. In fact, the unlimited number of new pairs (TISE, exact solution) can be generated in this way. The method is as well applicable to one-dimensional propagation of electromagnetic waves through inhomogeneous media with many important applications. Finally, section 7 briefly summarizes the obtained results.

## 2 TDHO Equation

The TDHO equation (TDHOE) reads

$$\ddot{x}(t) + \Omega^2(t) \cdot x(t) = 0\,, \tag{1}$$

where overdot represents differentiation with respect to time and time-dependent parameter $\Omega(t)$ is allowed to attain not only real but pure imaginary values as well (so that $\Omega^2$ may possibly become negative for some time intervals)[1]. Despite the apparent simplicity of the equation (1), there exists no known procedure of obtaining its general solution in *analytic form* for an arbitrarily *given* $\Omega^2(t)$. The powerful Wentzel–Kramers–Brillouin (WKB) method [2] of solving TDHOE is the most familiar example of approximating the true solution under some favorable conditions. Among various possible interpretations, the equation (1) can be thought of as the Newton's law of motion for a massive particle in a time-dependent one-dimensional harmonic potential. The time-dependent Hamiltonian corresponding to (1) is

$$H = \frac{p^2}{2M} + \tfrac{1}{2} M \cdot \Omega^2(t) \cdot x^2, \quad p = M \cdot \dot{x}\ , \tag{2}$$

where $M$ denotes the particle's mass and $x$ has the meaning of particle's location coordinate. Whenever $\Omega^2$ depends on time, the particle's energy is not a conserved quantity. However, for any time $t$, the Hamiltonian $H(t)$ equals the sum of the particle's kinetic and potential energies, which would be conserved at this level for all moments in the future provided that $\Omega^2(t)$ suddenly 'freezes' at the current value. Thus, one can think of $H(t)$ as the *instantaneous energy* of a mechanical oscillator (particle). From (1), we can substitute $\Omega^2(t) = -\frac{\ddot{x}}{x}$ and $p = M \cdot \dot{x}$ into (2) to obtain

$$H(t) = \frac{M}{2} \cdot [(\dot{x})^2 - x \cdot \ddot{x}] \equiv E_{TK}(t) \tag{3}$$

[1] There is nothing contradictory or 'unphysical' in letting $\Omega^2(t)$ to become negative for some time interval; it means only that the point $x = 0$ becomes a 'repelling point' for that period of time instead of being an 'attractive point' of motion.

The expression (3), up to a factor $M/2$, coincides with the so-called continuous-time *Teager-Kaiser (TK) energy operator* [3], [4] that has been used for tracking the energy of a source producing oscillations and for signal and speech AM-FM demodulation [5], [6]. From the derivation above, it is clear that formula (3) yields physically meaningful 'instantaneous energy' of a signal $x(t)$ only if $x(t)$ itself is a solution of the equation (1). In other words, formula (3) is applicable only for 'oscillating signals', for which $\Omega^2(t) = -\frac{\ddot{x}}{x}$ is a well-behaved function of time. In what follows, we will interchangeably refer to $E_{TK}(t)$ as *instantaneous energy* or *TK energy* (ignoring absence of constant factor $M/2$ in the standard definition [3]). Note that TK energy is distinct from the conventional designation in mathematical and signal processing literature that associates $x^2(t)$ with a measure of energy content in a signal.

Let us write the coordinate variable $x(t)$ in a polar form as follows

$$x(t) = \rho(t) \cdot \cos[\theta(t)], \tag{4}$$

where time-dependent amplitude $\rho$ and phase $\theta$ are introduced as new variables. Substituting (4) into (1), we obtain the system

$$\ddot{\rho} + \Omega^2(t)\rho - (\dot{\theta})^2 \rho = 0 \tag{5a}$$

$$\frac{1}{\rho} \cdot \frac{d}{dt}\left(\rho^2 \dot{\theta}\right) = 0 \tag{5b}$$

Equation (5b) can be rewritten as

$$M\, \rho^2 \dot{\theta} = L = constant, \tag{6}$$

which is just the angular momentum ($L$) conservation law for motion of a particle of mass $M$ in a planar field of central force. Substituting $\dot{\theta} = L/(M\rho^2)$ into (5a), we obtain the nonlinear *Ermakov-Pinney* differential equation [7], [8]:

$$\ddot{\rho} + \Omega^2(t)\rho - \frac{\widetilde{L}^2}{\rho^3} = 0 \tag{7}$$

where $\widetilde{L} \equiv L/M$ denotes the *angular momentum per unit of mass*.

The equations (6) and (7) determine the Hamiltonian dynamics of a particle in a plane in a time-dependent central-symmetric potential $V = \frac{1}{2} M\,\Omega^2(t)\rho^2$ using polar system of coordinates $(\rho,\theta)$. The Cartesian coordinates of the particle $x_1 = \rho\cos\theta,\, x_2 = \rho\sin\theta$, as functions of time, represent two linearly independent solutions of equation (1) with their Wronskian ($= x_1\dot{x}_2 - \dot{x}_1 x_2$) equal to $\widetilde{L}$. It is worth emphasizing that $L$ ($\widetilde{L}$) is *exact invariant* of motion which is conserved (unlike the adiabatic invariant) even when $\Omega(t)$ varies significantly within one period of oscillations. Also note that $L$ ($\widetilde{L}$) has a dimension of *action* (*action per unit of mass*). Moreover, constant $\widetilde{L}$ has a geometrical meaning of being twice the area traced by the radius-vector of a particle in the plane $(x_1, x_2)$ per unit of time.

Let us draw attention to the fact that one can obtain a pair of linearly independent solutions of equation (1) at once if, instead of looking for solution in the form (4), we allow $x(t)$ to be a complex-valued function: $x(t) = \rho(t)\cdot e^{i\theta(t)}$. Then real and imaginary components of $x(t)$ yield the abovementioned solutions $x_1(t)$ and $x_2(t)$, respectively.

The equation (7) can be interpreted as an equation governing radial motion of a particle in one-dimensional "effective" potential

$$V_\rho = \tfrac{1}{2} M\,\Omega^2(t)\rho^2 + \frac{L^2}{2M\rho^2} = \tfrac{1}{2} M\,\Omega^2(t)\rho^2 + \frac{M\widetilde{L}^2}{2\rho^2} \tag{8}$$

The corresponding "effective" Hamiltonian is

$$H_\rho = \frac{P_\rho^2}{2M} + \tfrac{1}{2} M\,\Omega^2(t)\rho^2 + \frac{M\widetilde{L}^2}{2\rho^2}\ ,\quad P_\rho = M\dot{\rho} \tag{9}$$

Note that the particle's trajectory never passes through the origin of coordinates provided that $\widetilde{L} \neq 0$ .

If the solution of equation (7) $\rho(t)$ is known, it can be used to find the phase variable $\theta(t)$ by integrating the equation (6):

$$\theta(t) = \theta(t_0) + \widetilde{L}\int_{t_0}^{t} \frac{d\tau}{\rho^2(\tau)} \tag{10}$$

The time derivative of phase variable can be associated with the *instantaneous angular frequency*:

$$\omega(t) \equiv \dot{\theta} = \frac{\widetilde{L}}{\rho^2(t)} \tag{11}$$

Let us now derive a very useful equation for the phase variable $\theta(t)$. Assume that $\dot{\theta} > 0$ for all times. Then, from (6), we have $\rho = \sqrt{\frac{\tilde{L}}{\dot{\theta}}}$ and $\ddot{\rho} = -\frac{\sqrt{\tilde{L}}}{2} \cdot \frac{\dddot{\theta} \cdot \dot{\theta} - \frac{3}{2}\ddot{\theta}^2}{\dot{\theta}^{5/2}}$. Substituting these results into the Ermakov-Pinney equation (7), we obtain the relationship

$$\Omega^2(t) = \dot{\theta}^2 + \tfrac{1}{2}\{\theta; t\} \tag{12}$$

where $\{\theta; t\}$ denotes the *Schwarzian derivative* (SD) of function $\theta(t)$ defined as

$$\{\theta; t\} = \frac{\dddot{\theta}}{\dot{\theta}} - \frac{3}{2}\left(\frac{\ddot{\theta}}{\dot{\theta}}\right)^2 \tag{13}$$

The Schwarzian derivative appears in many different branches of mathematics, from differential geometry of curves (where it is related to curvature) to the theory of conformal maps [9]. The remarkable feature of SD is that it is zero only for functions that are *linear fractional transformations* of the argument, i.e., $\{\theta; t\} = 0$ if and only if $\theta(t) = \widehat{T}_{abcd}(t) = \dfrac{a \cdot t + b}{c \cdot t + d}$, where *a*, *b*, *c*, and *d* are constants with the determinant $ad - bc \neq 0$. For equation (12) this means that, if $\theta(t)$ is not a linear fractional transformation, $\dot{\theta}$ cannot follow $\Omega(t)$ exactly without some offset. Another interesting feature of SD is its invariance under linear fractional transformation of the function it operates upon, i.e., $\{\widehat{T}_{abcd}(f); t\} = \{f; t\}$.

## 3 Method of Finding Pairs (TDHOE, Exact Solution)

As mentioned earlier, at present no general method of solving TDHOE is known for an arbitrary given function $\Omega^2(t)$. Only a small number of pairs (TDHOE, exact solution) are identified. The most celebrated example is the set of TDHOEs with $\Omega_n^2(t) = 2n + 1 - t^2$ ($n$ = 0, 1, 2, …) for which the set of corresponding exact solutions in $L^2$ space consists from the Hermite functions

$$H_n(t) = \frac{1}{\pi^{1/4}\sqrt{2^n n!}} \exp\left(-\tfrac{t^2}{2}\right) \cdot h_n(t), \tag{14}$$

where $h_n(t)$ are the Hermite polynomials:

$$h_n(t) = (-1)^n e^{t^2} \tfrac{d^n}{dt^n}(e^{-t^2}) \tag{15}$$

Due to linearity of TDHOE, multiplication of $H_n(t)$ by an arbitrary constant factor gives also a solution. The factors in (16) are chosen such that the normalization conditions $\int_{-\infty}^{+\infty} H_n^2(t)dt = 1$ are satisfied. The motive for this normalization can be understood if, instead of time, we think of $t$ as representing a dimensionless spatial coordinate. Then the TDHOE with $\Omega_n^2(t) = 2n + 1 - t^2$ formally coincides with the appropriately normalized TISE for a quantum harmonic oscillator and the Hermite functions play the role of its orthonormal eigenfunctions. This interesting link between TDHO and TISE will be generalized and discussed in detail later in section 4.

This section is concerned with an explicit method for finding pairs (TDHOE, exact solution). Although this method is not general enough to produce all possible such pairs, it can generate many previously unknown pairs of high theoretical and practical value.

A naive approach to generating a pair (TDHOE, exact solution) would be to select some smooth function $x(t)$ and then calculate the corresponding $\Omega^2(t)$ from the equation (1): $\Omega^2(t) = -(\ddot{x}/x)$. However, this method most often produces discontinuities in parameter $\Omega(t)$ and therefore is of little interest. By contrast, the method described below *always* produces smooth exact solutions of TDHOEs with continuous parameter $\Omega(t)$:

1. Select a phase variable $\theta(t)$ as some known continuous function of time with continuous time derivatives up to a third order at least and positive first order derivative ( $\dot{\theta}(t) > 0$ ) for all times.

2. Calculate $\Omega^2(t)$ from the equation (12). Note that, since $\theta(t)$ is known, (12) is not a non-linear differential equation to be solved but simply a formula to express $\Omega^2(t)$ via $\dot{\theta}(t)$ and $\{\theta;t\}$ .

3. Substitute $\Omega^2(t)$ obtained in step 2 into equation (1). The result is a particular form of TDHOE, the first element of the sought-after pair.

4. Calculate $x(t) = \sqrt{\frac{\widetilde{L}}{\dot{\theta}(t)}} \cos[\theta(t)]$, where $\widetilde{L}$ is an arbitrary positive constant. By construction, $x(t)$ is an *exact solution* of the TDHOE found in the step 3 and thus is a second element of the sought-after pair.

## 4 Relationship between TDHO and Non-Relativistic Quantum Mechanics in One Dimension

This section reveals the surprisingly close relationship between one-dimensional non-relativistic quantum mechanics and classical physics of TDHO. We start from the general TISE

$$\frac{d^2\psi}{d\,x^2} + \frac{2m}{\hbar^2}[E - U(x)]\psi = 0 \tag{16}$$

and rewrite the potential $U(x)$ in the form:

$$U(x) = U_0 \cdot \overline{u}_1(\tfrac{x}{a}) \tag{17}$$

where $U_0$ has the dimension of energy and represents the characteristic energy scale of the problem, and $\overline{u}_1$ denotes a dimensionless function of a dimensionless argument x/*a* with *a* being some characteristic length. Buckingham's pi-theorem [10] guaranties that such representation is always possible for any physically meaningful potential function. Then we introduce the characteristic angular frequency $\omega$ via the relation:

$$\omega = \frac{2U_0}{\hbar} \tag{18}$$

and the new length scale $x_0$ as

$$x_0 = \left(\frac{\hbar}{m\omega}\right)^{\frac{1}{2}} = \hbar(2mU_0)^{-\frac{1}{2}} \tag{19}$$

Now we can go back to (17) and rewrite it as

$$U(x) = U_0 \cdot \overline{u}_1(\tfrac{x}{x_0} \cdot \tfrac{x_0}{a}) = U_0 \cdot \overline{u}(\tfrac{x}{x_0}) \tag{20}$$

where we introduced a new dimensionless function $\overline{u}$ absorbing the parameter $x_0 / a$ inside it. Admitting new variables $\overline{x} = x / x_0, \overline{E} = E / U_0$, and $\overline{\psi} = x_0^{\frac{1}{2}} \cdot \psi$ we can finally write the dimensionless TISE:

$$\frac{d^2\overline{\psi}}{d\overline{x}^2} + [\overline{E} - \overline{u}(\overline{x})]\overline{\psi} = 0 \tag{21}$$

On the other hand, TDHOE (1) can also be expressed in dimensionless form:

$$\frac{d^2\overline{X}}{d\overline{t}^2} + \overline{\Omega}^2(\overline{t})\overline{X} = 0 \tag{22}$$

where the new variables are defined as:

$\overline{t} = \nu \cdot t;\quad \overline{X} = x / r_0;\quad \overline{\Omega}^2 = \Omega^2 / \nu^2;\quad \overline{E}_{TK} = E_{TK} / V_0\quad$ with $\nu = \frac{2V_0}{L};\ r_0 = \left(\frac{L}{M\nu}\right)^{\frac{1}{2}}$.

As previously, constant $L$ has a dimension of action (or angular momentum) and is analogous to the reduced Planck constant $\hbar$, and $V_0$ sets the energy scale. Now, if we select

$$\overline{\Omega}^2(\overline{t}) \equiv \overline{E} - \overline{u}(\overline{t})\ , \tag{23}$$

the equations (21) and (22) become literarily equivalent and hence their solutions under the appropriate boundary conditions completely match each other: $\overline{\psi}(\overline{x}) \equiv \overline{X}(\overline{t})$. Hence the dimensionless wavefunction – solution of the TISE (21) – can be simultaneously thought of as a dimensionless trajectory of a classical particle moving in the corresponding time-dependent harmonic potential

$$V = V_0 \cdot [\overline{E} - \overline{u}(\overline{t})] \cdot \overline{X}^2\ , \tag{24}$$

thus revealing a surprising link between classical and quantum physics in one dimension. In what follows, we will refer to the classical and quantum particles related via such link as *analogs*. Notice that, in case of a quantum particle, its energy is given by $\bar{E}U_0$, while the instantaneous energy of its classical analog reads

$$E_{TK}(\bar{t}) = V_0 \cdot [\dot{\bar{X}}^2(\bar{t}) + (\bar{E} - \bar{u}(\bar{t}))\bar{X}^2(\bar{t})] \tag{25}$$

Table 1 summarizes the isomorphism between TDHOE and TISE. The quantities in the same row are analogous.

**Table 1:** Isomorphism between TDHOE and TISE in one dimension

| **TDHOE** | | **TISE** | |
|---|---|---|---|
| ***Variable*** | ***Meaning*** | ***Variable*** | ***Meaning*** |
| $\bar{t}$ | time | $\bar{x}$ | spatial coordinate |
| $\bar{X}$ | spatial coordinate | $\bar{\psi}$ | wave function |
| $L$ | angular momentum | $\hbar$ | reduced Planck's constant |
| $\bar{E}$ | parameter in time-dependent harmonic potential | $\bar{E}$ | particle's energy |
| $M$ | mass | $m$ | mass |

## 5 'Range Relations' in Classical Mechanics?

The link between classical and quantum physics exposed in the previous section provokes an interesting question, namely whether there exist some intrinsic features of TDHO that make it uniquely positioned to serve as a link between classical and quantum domains. In this section, we will investigate the reaction of TDHO to sudden parametric perturbations using the method for finding pairs (TDHOE, exact solution) presented in section 3. We will observe the behavior of a classical (macroscopic) TDHO that can be characterized by the 'range relations' analogous to the uncertainty relations of quantum mechanics. The fact that only *exact solutions* of TDHOE will be used in our study significantly simplifies interpretation of the obtained results, excluding the possibility of artifacts caused by the approximations.

Specifically, consider a one-dimensional TDHO which phase is (parametrically) perturbed according to the following law

$$\theta(t) = \theta_0 + \omega_0 t + a \cdot \exp[-\tfrac{t^2}{T^2}], \quad a > 0 \tag{26}$$

where $\theta_0$ is the initial phase, $\omega_0$ - the unperturbed angular frequency, $a$ - the amplitude of phase perturbation, and $T$ - the characteristic time of perturbation. The parameters $a$, $\omega_0$, and $T$ are related via the inequality which assures that $\dot{\theta} > 0$ for all times:

$$T > T_{\min} = \frac{a}{\omega_0}\sqrt{\frac{2}{e}} \tag{27}$$

For fixed values of parameters $a$, $\omega_0$ and different values of $\theta_0\,(0 \le \theta_0 < 2\pi)$ and $T$ that satisfy (27), one can find the corresponding TDHOEs and their *exact* solutions $x(t) = \rho(t)\cos[\theta(t)]$ along with the associated instantaneous energies $E_{TK}(t)$ and momentums $p(t) = M\,\dot{x}(t)$. The unperturbed (reference) motion of the oscillator ($a$ = 0) is given by

$$x_{ref}(t) = \rho_0 \cos[\theta_0 + \omega_0 t] \tag{28}$$

and hence the value of the *exact invariant L* can be found as

$$L = M\rho^2(t)\dot{\theta}(t) = M\rho_0^2\omega_0 \tag{29}$$

*L* remains unchanged during all times even after we switch the (parametric) perturbation on. Let us fix the values of several uncritical parameters as follows

$M$ = 0.1 kg; $\rho_0 = 0.01\,m$; $\omega_0 = 2\pi \cdot 5$ rad/s; $a$ = 1 rad.

One of the typical results obtained is shown in Figure 1. Figure 1B presents the behavior of $\Omega^2(t)$, which parametrically generates the perturbation of phase $\theta(t)$ shown in Figure 1A (note that $\Omega^2(t)$ and $\theta(t)$ were calculated in the reverse order, see section 3). From Figure 1C, we observe the behavior of the instantaneous energy $E_{TK}(t)$ exhibiting the spike and the fall to the negative area during time of perturbation. The quantity

$$\delta E_{TK} \equiv \max_{t\in(-\infty,+\infty)} E_{TK}(t) - \min_{t\in(-\infty,+\infty)} E_{TK}(t) \tag{30}$$

is the *energy range* of the oscillator during perturbation. Figure 1D shows the perturbed position of the oscillator. Similarly to the previous case, we define the *position range* and the *momentum range* of the oscillator respectively as

$$\delta X \equiv \max_{t\in(-\infty,+\infty)} [x(t) - x_{ref}(t)] - \min_{t\in(-\infty,+\infty)} [x(t) - x_{ref}(t)] \tag{31}$$

and

$$\delta P \equiv \max_{t\in(-\infty,+\infty)} [p(t) - p_{ref}(t)] - \min_{t\in(-\infty,+\infty)} [p(t) - p_{ref}(t)] \tag{32}$$

Let us investigate the conjecture that there exist 'range relations' for a macroscopic TDHO that can be regarded as analogs of the uncertainty relations of quantum mechanics, with classical energy, position, and momentum ranges playing the roles of quantum energy, position, and momentum uncertainties, respectively. Certainly, the interpretations of the 'range relations' (if they exist) in a macroscopic case would be significantly different from the present interpretation of the uncertainty relations of quantum mechanics. Nevertheless their mere existence might shed new light on the relationship between classical and quantum physics in one dimension.

First, let us introduce the classical ranges averaged over uniformly distributed initial phase $\theta_0$ as follows:

$$\Delta E_{TK} \equiv \left\langle \delta E_{TK}(\theta_0) \right\rangle_{\theta_0} \ ; \ \Delta X \equiv \left\langle \delta X(\theta_0) \right\rangle_{\theta_0} \ ; \ \Delta P \equiv \left\langle \delta P(\theta_0) \right\rangle_{\theta_0} \tag{33}$$

Then the conjectured macroscopic 'range relations' for a TDHO read

***Time-Energy***: $\Delta t \cdot \Delta E_{TK} \geq \frac{1}{2} L$ (34)

***Position-Momentum***: $\Delta X \cdot \Delta P \geq \frac{1}{2} L$ (35)

One can put forth the hypothesis that these range relations are fulfilled for an *arbitrary* parametric perturbation of a classical TDHO with characteristic perturbation duration $\Delta t$. The comprehensive numerical verification and (possibly) formal proof of this hypothesis will be the subject of future research.

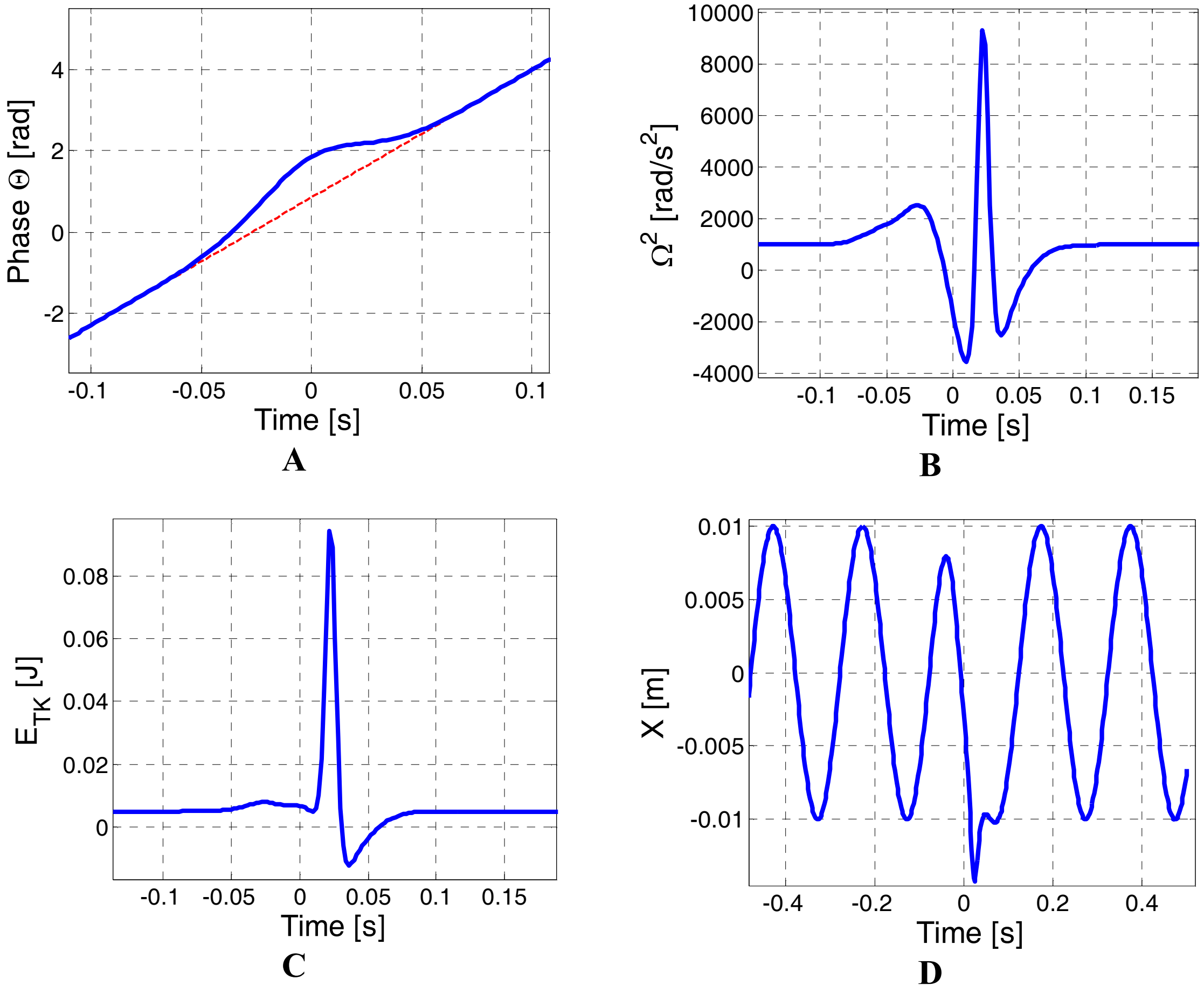

**Figure 1:** (A) Typical (parametric) perturbation of phase $\theta(t)$; (B) parameter $\Omega^2(t)$; (C) the corresponding instantaneous energy $E_{TK}(t)$; (D) the corresponding perturbed position (coordinate) of TDHO. See text for details.

Meanwhile here I present the numerical evidence of the validity of the conjectured relations (34) and (35) for the particular type of perturbations given by (26). Obviously, one should set $\Delta t = T$ in this case. The typical results of calculations based on *exact solutions* of the relevant TDHOE are shown in Figure 2. The initial phase varied from 0 to $2\pi$ with increment $\pi/180$ rad (= 1°) for every value of the perturbation interval *T*, which in turn varied from $1.2 \cdot T_{\min}$ to $4 \cdot T_{\min}$ with increment $0.1 \cdot T_{\min}$. The dots represent values of $\frac{2}{L}\delta E_{TK} T$ and $\frac{2}{L}\delta X \delta P$ for various initial phases $\theta_0$ and stars connected with a line stand for the values of $\frac{2}{L}\Delta E_{TK} T$ and $\frac{2}{L}\Delta X \Delta P$. We observe excellent agreement with the hypothetical inequalities (34) and (35).

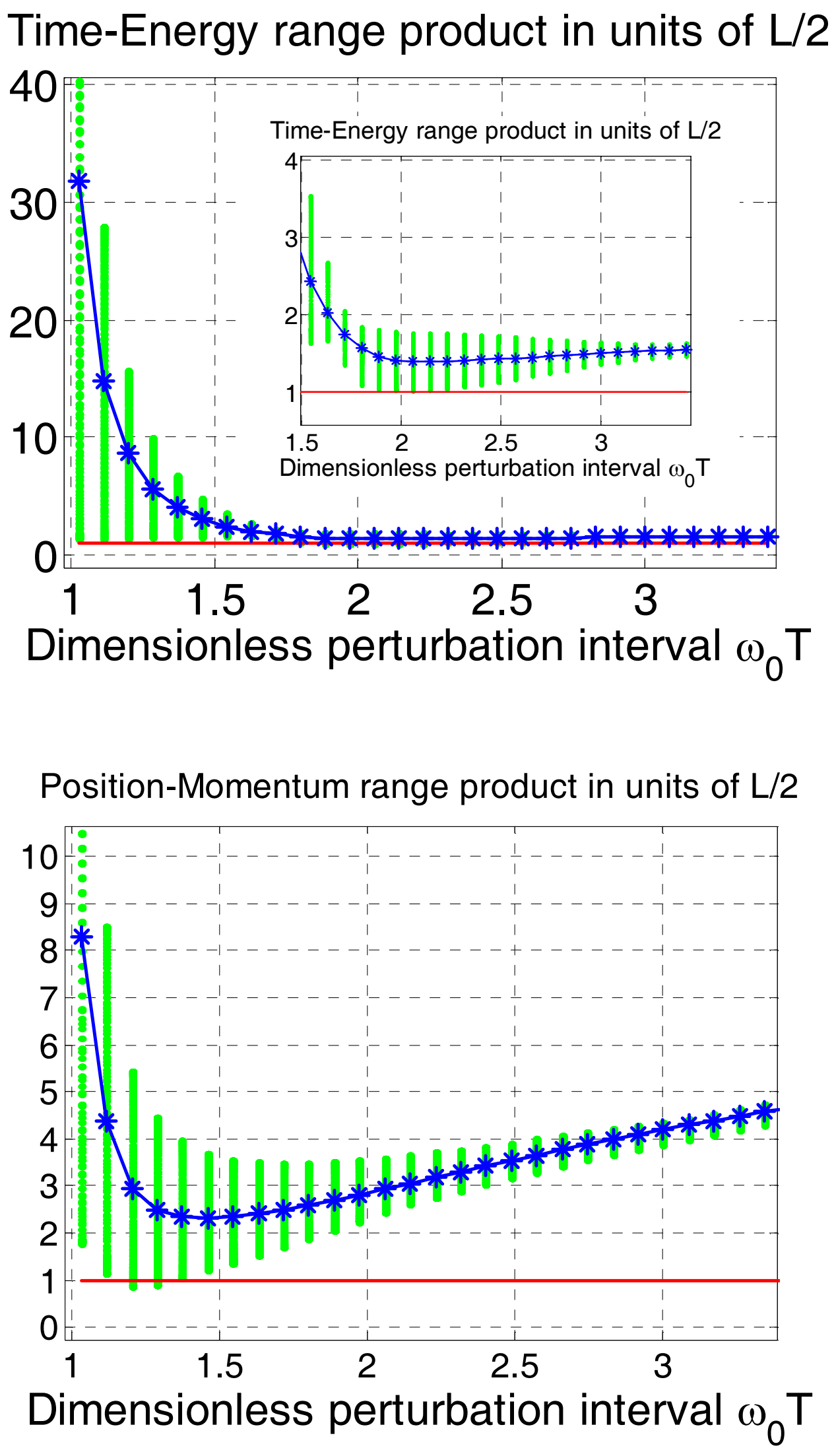

**Figure 2:** Experimental support for validity of *Time-Energy* and *Position-Momentum* range relations for a macroscopic TDHO. See text for details.

## 7 Tunnel Effect: New Exact Solutions and Classical Outlook

Quantum tunneling is one of the most studied problems in quantum mechanics being equally of fundamental and practical importance. In this section, we first apply the method introduced earlier to obtain an explicit formula generating a wide class of exact solutions of the TISE describing one-dimensional tunneling through previously uninvestigated potential barriers, and then discuss the interpretation of these solutions in terms of a classical TDHO.

Consider a one-dimensional tunneling problem described by TISE in the dimensionless form (21). Assume that quantum particles impinging on a potential barrier (which is located around $\bar{x} = 0$ and is vanishing in the limits $\bar{x} \to \pm\infty$) from the left possess identical dimensionless energy $\bar{E} = 1$ (i.e., we admit the kinetic energy of a particle far to the left from the barrier as a characteristic energy $U_0$). In the limit $\bar{x} \to -\infty$, the wavefunction can be decomposed into a superposition of impinging and reflected waves:

$$\bar{\psi} = A(e^{i\cdot\bar{x}} + \sigma_0 e^{-i\cdot\bar{x}}), \tag{36}$$

where $A$ denotes normalization constant and $\sigma_0$ is related to the *reflection coefficient R* via $R = |\sigma_0|^2$ ($0 \le R < 1$). Both $A$ and $\sigma_0$ are complex-valued in general case. Here we will consider only real-valued $\sigma_0$, thus notably simplifying the resulted formulae. The generalization of the method introduced below for a complex-valued $\sigma_0$ is straightforward but bulky and will be left outside the scope of this paper. For real-valued $\sigma_0$, formula (36) can be rewritten in the polar form as

$$\bar{\psi} = |A| \cdot \sqrt{1 + \sigma_0^2 + 2\sigma_0 \cos(2\bar{x})} \cdot e^{i\theta(\bar{x},\sigma_0)} \tag{37}$$

with

$$\theta(\bar{x},\sigma_0) = \theta_0 + \arctan\left[\frac{1-\sigma_0}{1+\sigma_0} \cdot \tan(\bar{x})\right] \tag{38}$$

where phase constant $\theta_0 \in [0, 2\pi[$ originates from representation of $A$ in the polar form. Far to the right from the barrier ($\bar{x} \to +\infty$), the wavefunction must be proportional to $e^{i\cdot\bar{x}}$ describing a flow of quantum particles transmitted through the barrier. Let us introduce a continuous function $\sigma(\bar{x})$ such that

$$\sigma(\bar{x}) = \begin{cases} \sigma_0 \equiv \sqrt{R}, & \bar{x} \to -\infty \\ 0, & \bar{x} \to +\infty \end{cases} \tag{39}$$

Then the phase function defined as

$$\theta[\bar{x},\sigma(\bar{x}),\Delta\theta,D] = \theta_0 + \arctan\left[\frac{1-\sigma(\bar{x})}{1+\sigma(\bar{x})} \cdot \tan(\bar{x})\right] + \Delta\theta \cdot \frac{1 + \tanh(\bar{x}/D)}{2} \tag{40}$$

satisfies the required boundary conditions

$$\theta = \begin{cases} \theta_0 + \arctan\left[\dfrac{1-\sigma_0}{1+\sigma_0} \cdot \tan(\bar{x})\right], & \bar{x} \to -\infty \\ \theta_0 + \bar{x} + \Delta\theta, \quad \bar{x} \to +\infty \end{cases} \tag{41}$$

and hence may serve as an input to the algorithm presented in section 3. In formulae (40) and (41), $\Delta\theta$ denotes the phase shift after passing through the barrier (transmission phase shift) and $D$ is a dimensionless parameter controlling the relative width where this phase shift takes place. The functional dependency (40) is not the most general one but it is broad enough to offer a very wide class of exact solutions of the tunneling problem as will become evident shortly. One of possible choices for $\sigma(\bar{x})$ is given by

$$\sigma(\bar{x}) = \sigma_0 \cdot \frac{1 - \tanh(\bar{x}/d)}{2} + \eta(\bar{x}), \tag{42}$$

where $d$ denotes some (dimensionless) parameter ($d \sim 1$) and $\eta(\bar{x})$ is a small arbitrary function that vanishes in the limits $\bar{x} \to \pm\infty$ and is also constrained by the requirement $\frac{d\theta}{dx} > 0$. In case one wishes to consider the *reflectionless* barrier ($\sigma_0 = 0$) with vanishing phase shift ($\Delta\theta = 0$), function $\eta(\bar{x})$ must not be identically zero in order to give a non-trivial barrier shape. The solution of the equation (21) can be presented in the form $\bar{\psi}(\bar{x}) = \dfrac{B}{\sqrt{\theta'(\bar{x})}} \cdot e^{i\theta(\bar{x})}$, where $\theta' \equiv \frac{d\theta}{dx}$ and $B$ is a constant determined from the boundary conditions (37) and (41): $B = |A| \cdot \sqrt{1-R}$.
Thus we have

$$\bar{\psi}(\bar{x}) = |A| \cdot \sqrt{\frac{1-R}{\theta'(\bar{x})}} \cdot e^{i\theta(\bar{x})} \tag{43}$$

Notice that now $\bar{\psi}(\bar{x})$ is a complex-valued function, with phase $\theta$ given by (40). Recalling that $\bar{E} = 1$, we then obtain the potential barrier profile corresponding to the exact solution (43) via the relationships (12) and (23):

$$\bar{u}(\bar{x}) = 1 - (\theta')^2 - \tfrac{1}{2}\{\theta; \bar{x}\} \tag{44}$$

Thus, equations (40), (42), (43), and (44) together provide solution to the quantum tunneling problem with the pre-selected reflection coefficient $R$ ($R = |\sigma_0|^2$) and transmission phase shift $\Delta\theta$. The choice of parameters $d$ and $D$ as well as function $\eta(\bar{x})$ ($\eta(\bar{x}) \to 0$ as $\bar{x} \to \pm\infty$) is flexible, with the single restraining condition: $\frac{d\theta}{dx} > 0$. Observe that both the potential barrier profile $\bar{u}(\bar{x})$ and the wavefunction $\bar{\psi}(\bar{x})$ can be written in analytic form provided that phase function $\theta(\bar{x})$ is known analytically and is sufficiently smooth. In our approach, no connection formulas are necessary since the expression (43) defines the wavefunction on a whole domain $-\infty < \bar{x} < +\infty$, before, inside, and after the barrier potential. Notice also that, since formula (44) contains only the derivatives of the phase function, the potential $\bar{u}(\bar{x})$ is independent from the choice of the phase constant $\theta_0$. The same holds true for the relative probability density $|\bar{\psi}(\bar{x})|^2$ as is obvious from (43).

Many previously unknown *exact solutions* of a quantum tunneling problem in one dimension can be obtained with this method. For the illustration purposes, three examples are given below.

***Example 1.*** $R = 0.2$ $(Tr \equiv 1 - R = 0.8)$; $\Delta\theta = 0$; $|A| = 1$; $d = 2$; $\eta(\bar{x}) \equiv 0$. The phase function $\theta(\bar{x})$, probability density $|\bar{\psi}(\bar{x})|^2$, and the barrier potential $\bar{u}(\bar{x})$ are shown in Figure 3.

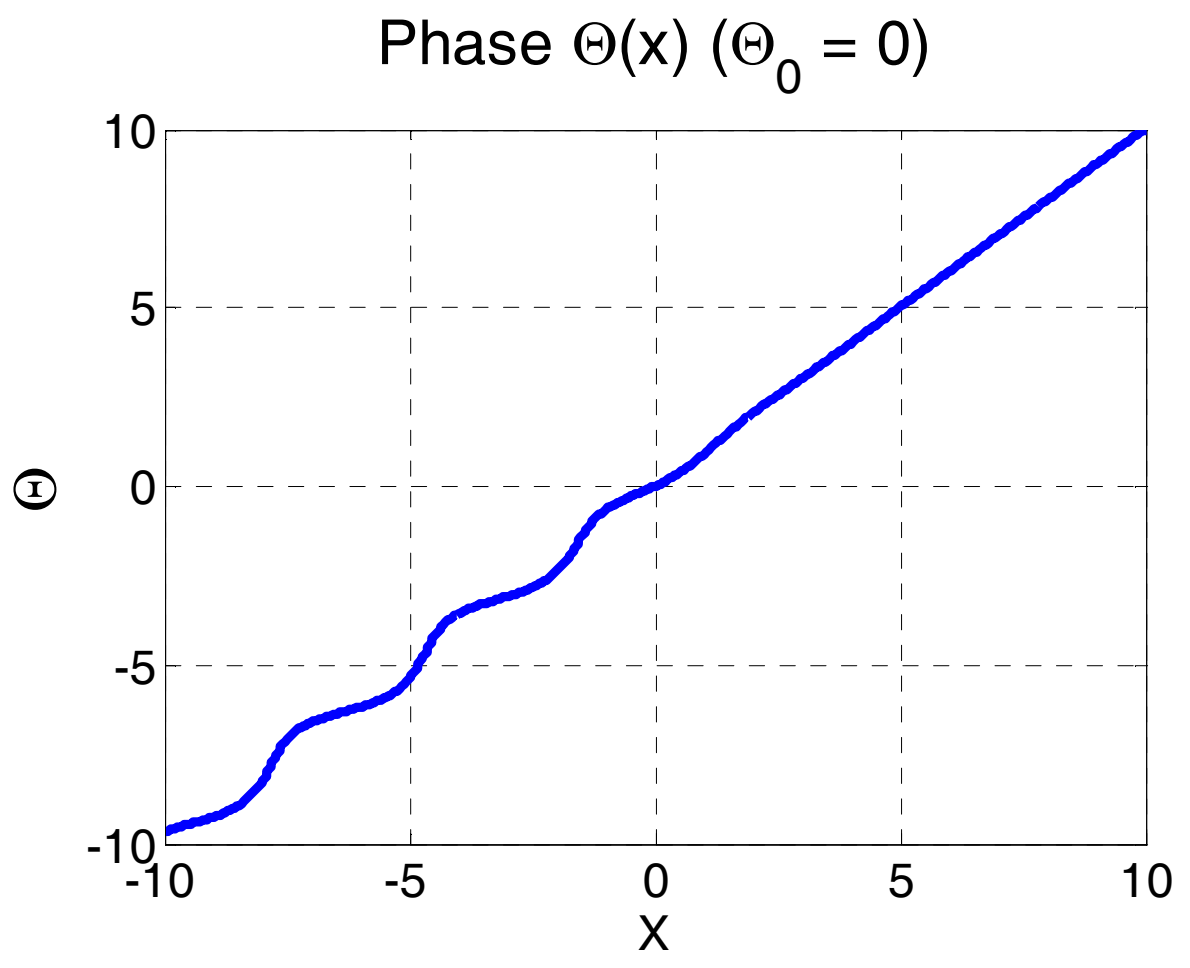

**(A)**

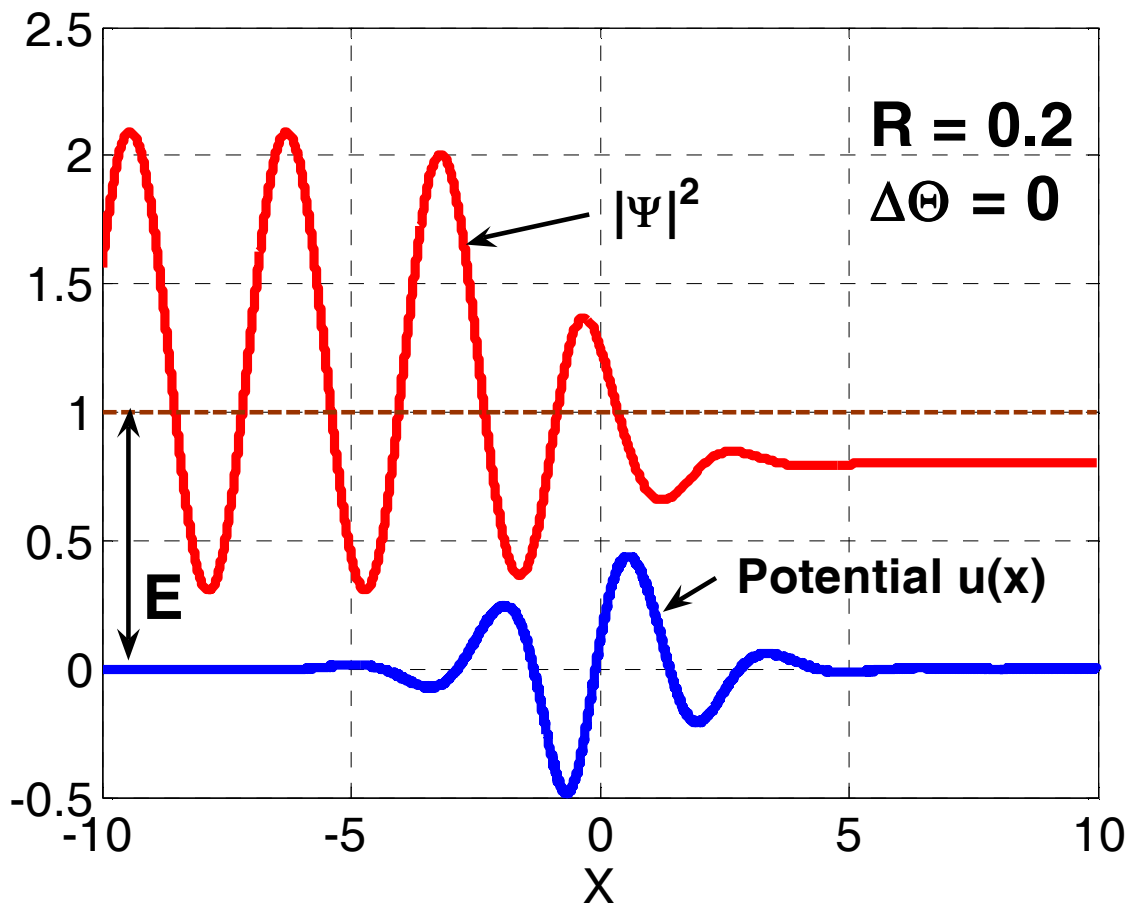

**(B)**

**Figure 3:** Results for Example 1. (A) Phase function $\theta(\bar{x})$; (B) probability density $|\bar{\psi}(\bar{x})|^2$, and the barrier potential $\bar{u}(\bar{x})$. This example describes particle current with over-barrier reflection probability $R = 0.2$ and transmission phase shift $\Delta\theta = 0$.

***Example 2.*** $R = 0\ (Tr = 1);\ \Delta\theta = -\pi;\ |A| = 1;\ D = 2;\ \eta(\bar{x}) \equiv 0$. The phase function $\theta(\bar{x})$, probability density $|\bar{\psi}(\bar{x})|^2$, and the barrier potential $\bar{u}(\bar{x})$ are shown in Figure 4.

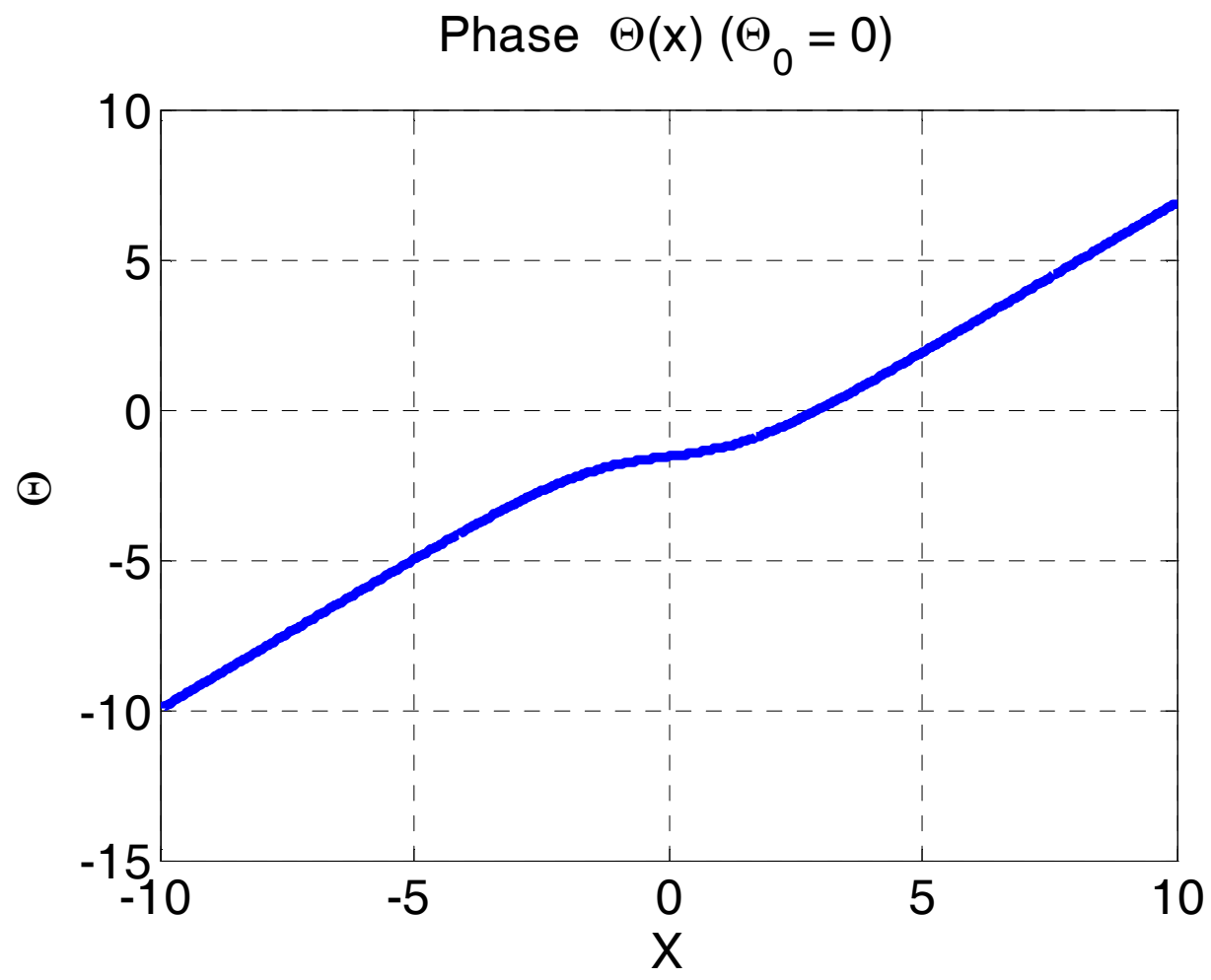

**(A)**

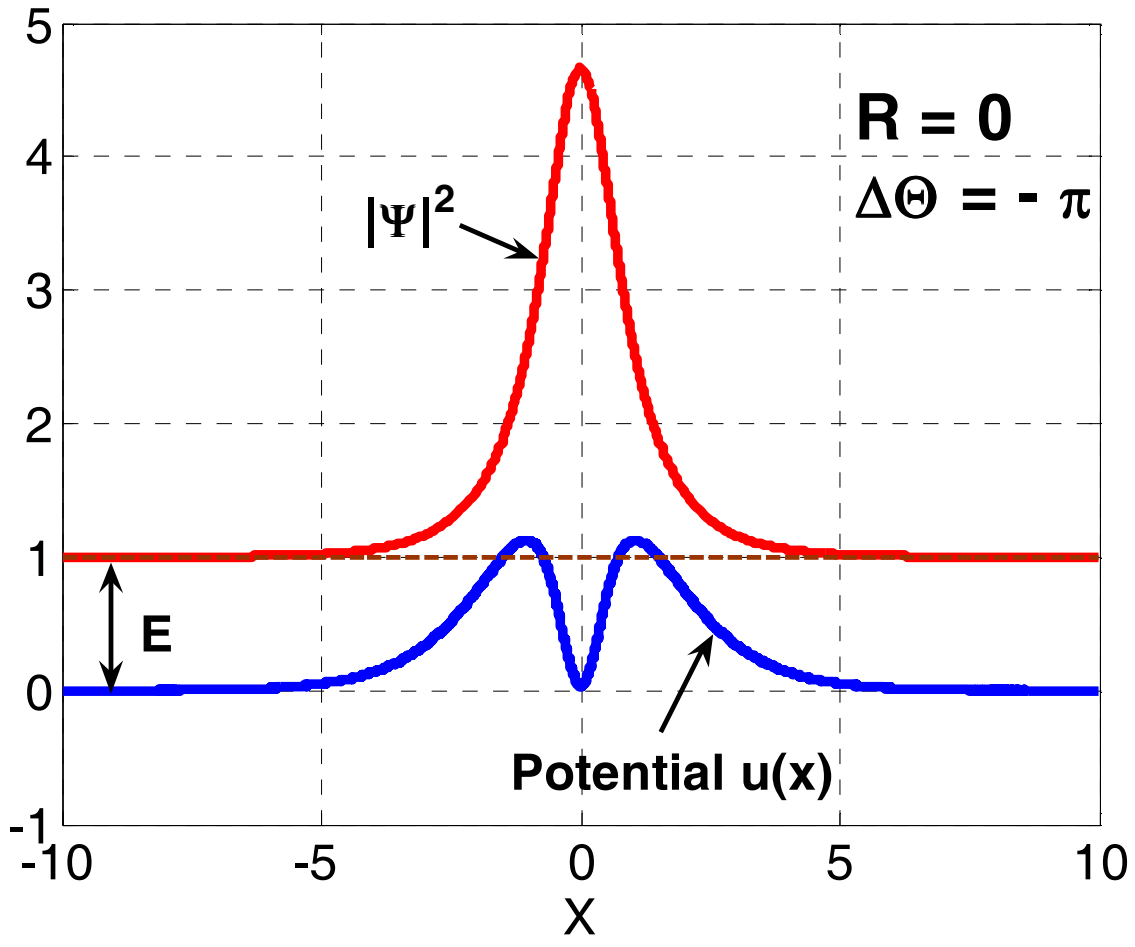

**(B)**

**Figure 4:** Results for Example 2. (A) Phase function $\theta(\bar{x})$; (B) probability density $|\bar{\psi}(\bar{x})|^2$, and the barrier potential $\bar{u}(\bar{x})$. This example describes reflectionless ($R$ = 0) tunneling with transmission phase shift $\Delta\theta = -\pi$.

***Example 3.***

$R = 0.05\ (Tr = 0.95);\ \Delta\theta = -\pi/2;\ |A| = 1;\ d = 1.25; D = 1.5;$

$$\eta(\bar{x}) = 0.1 \cdot \exp\left[-\frac{(\bar{x}+1)^2}{2d^2}\right] - 0.15 \cdot \exp\left[-\frac{(\bar{x}-1)^2}{2d^2}\right]$$

The phase function $\theta(\bar{x})$, probability density $|\bar{\psi}(\bar{x})|^2$, and the barrier potential $\bar{u}(\bar{x})$ are shown in Figure 5.

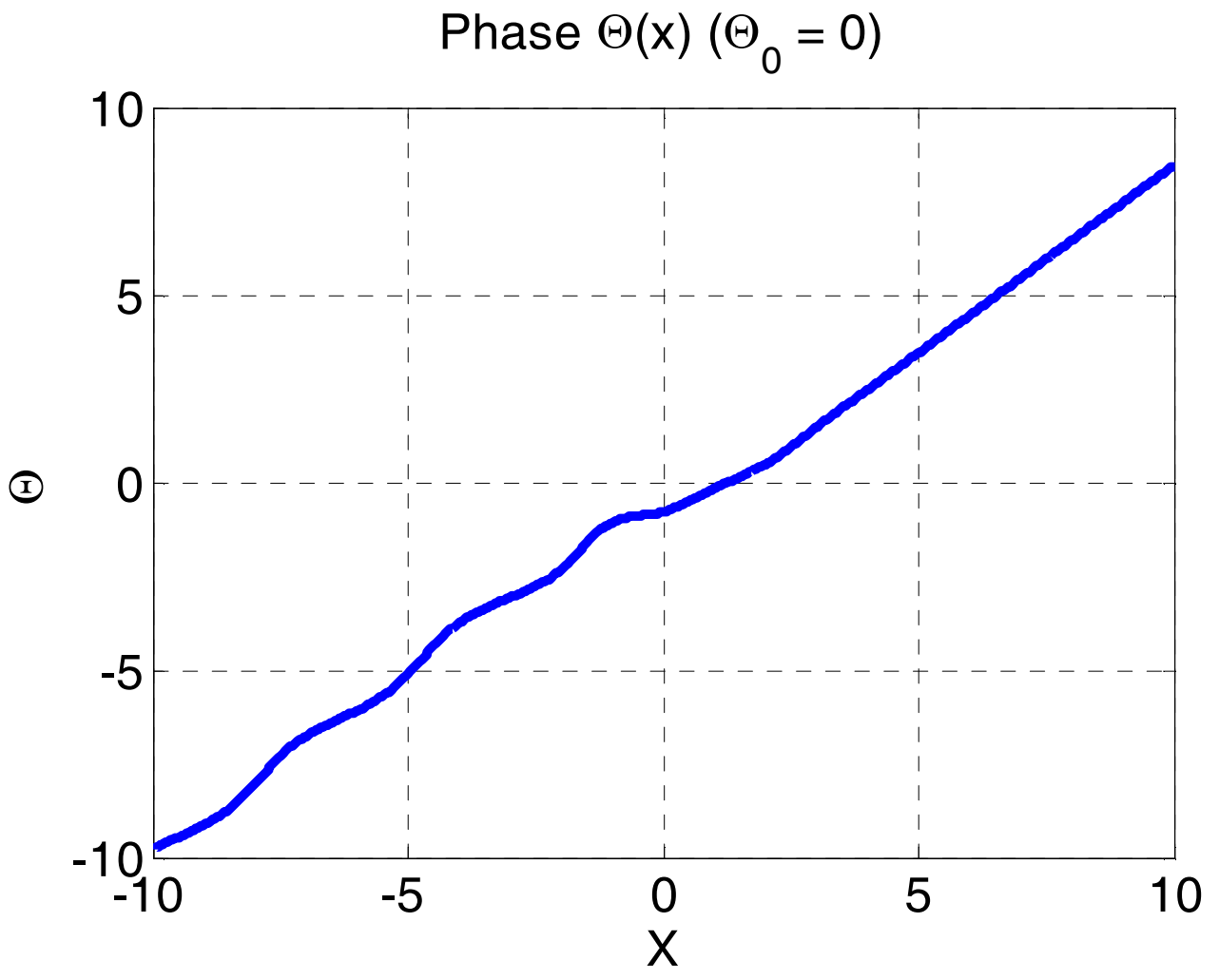

**(A)**

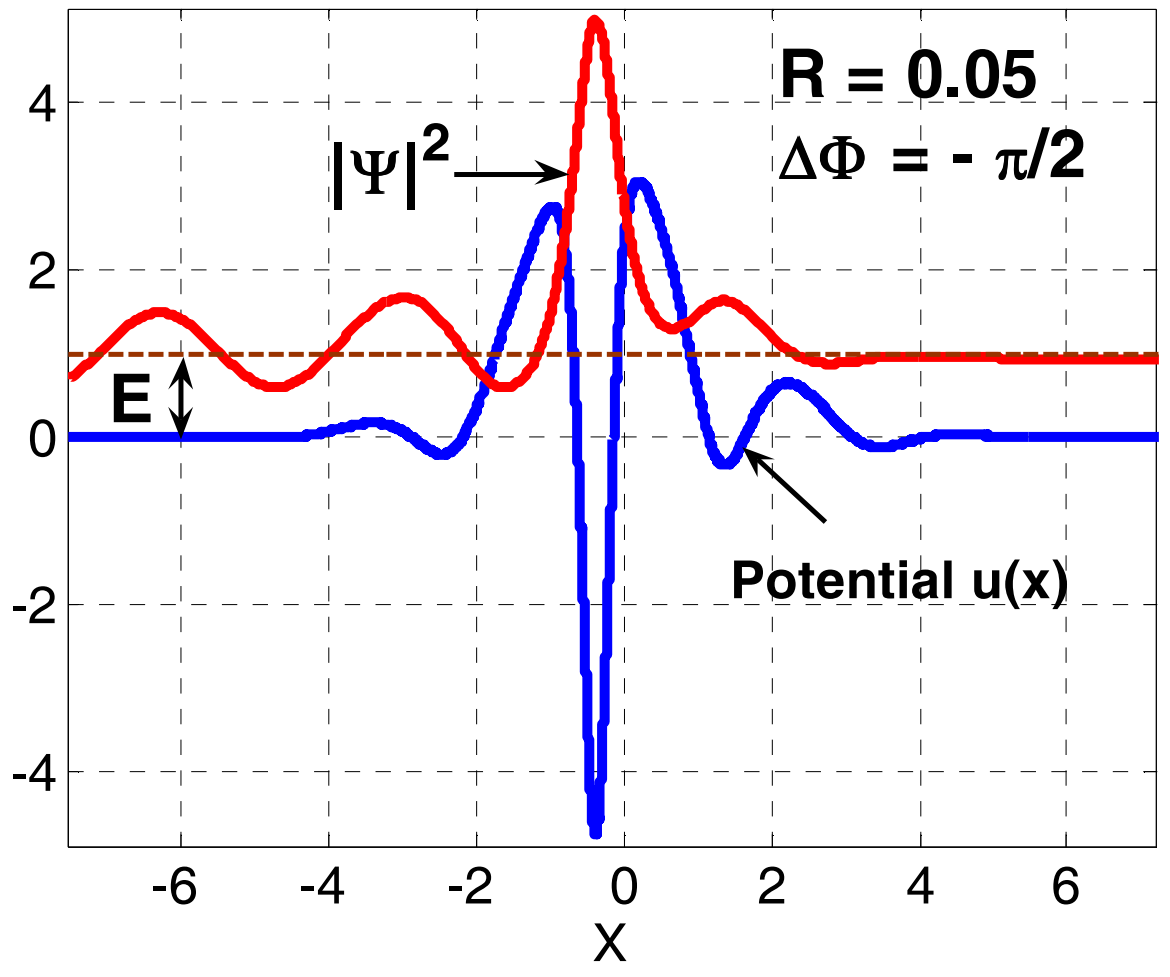

**(B)**

**Figure 5:** Results for Example 3. (A) Phase function $\theta(\bar{x})$; (B) probability density $|\bar{\psi}(\bar{x})|^2$, and the barrier potential $\bar{u}(\bar{x})$. This example describes tunneling with reflection coefficient $R$ = 0.05 and transmission phase shift $\Delta\theta = -\pi/2$.

In the light of the connection between one-dimensional quantum and classical physics revealed in section 4, let us now look at the typical classically-forbidden potential barrier penetration by a quantum particle from the classical standpoint. From formula (23), we can find the parameter $\bar{\Omega}^2(\bar{t})$ that corresponds to the classical counterpart of the tunneling problem. Figure 6 illustrates this correspondence. Notice that the classical turning points match the points where $\bar{\Omega}^2(\bar{t})$ vanishes. The associated time-dependent harmonic potential is presented in Figure 7.

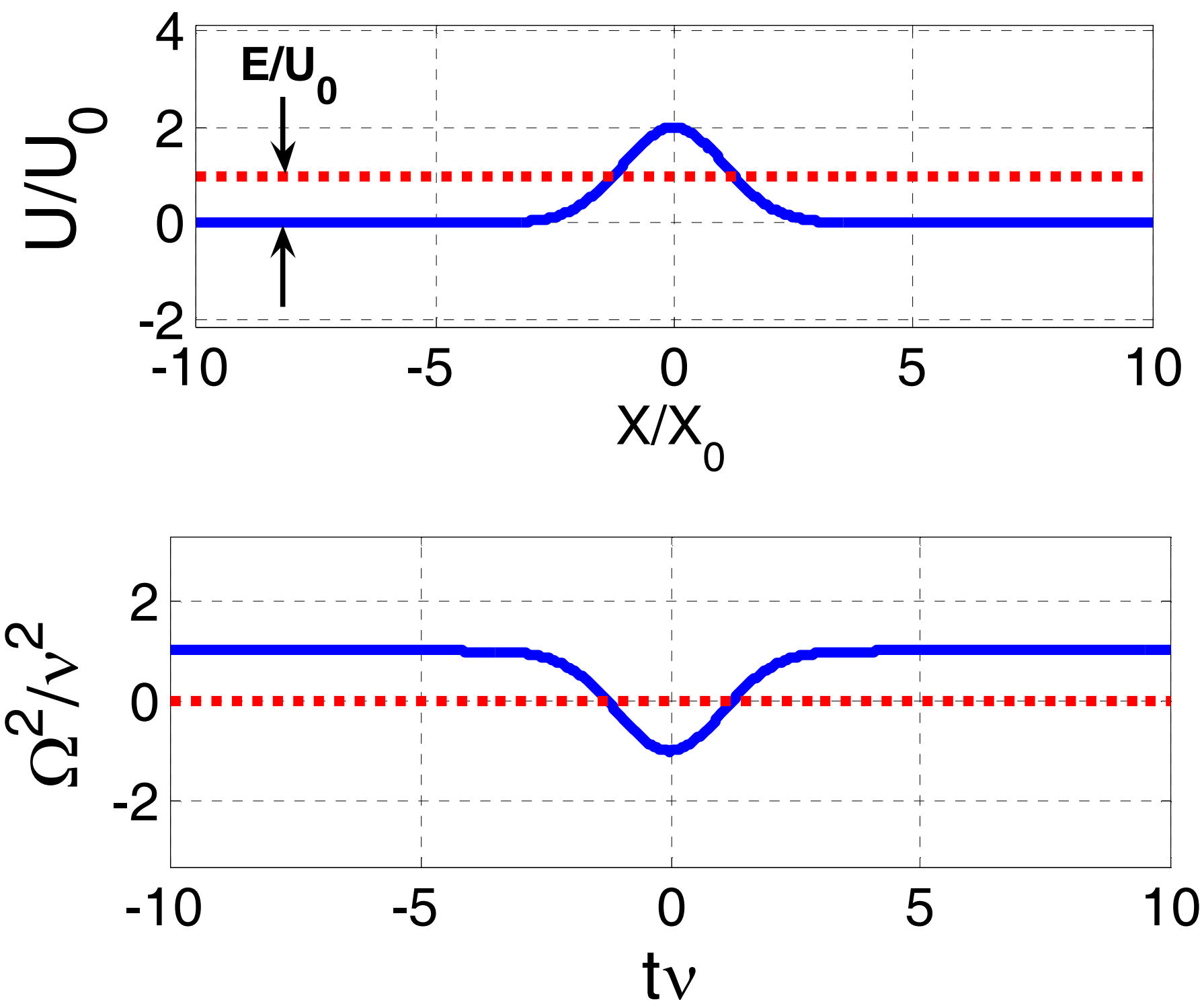

**Figure 6:** Correspondence between shape of the potential barrier in quantum tunneling problem and the associated parametric perturbation of TDHO.

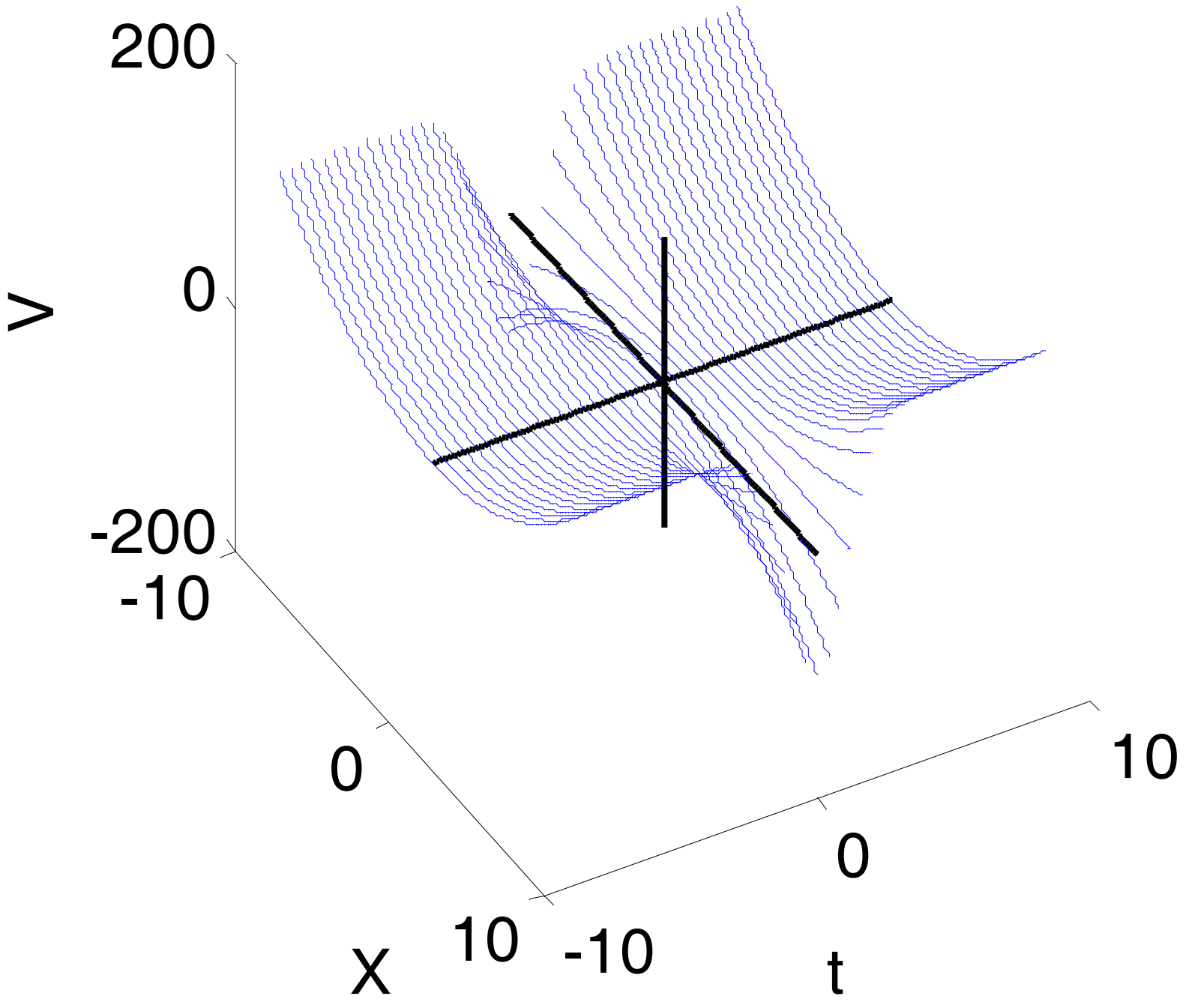

**Figure 7:** Time-dependent harmonic potential for analog of a typical quantum tunneling problem presented in Fig. 6

Two linearly independent real-valued solutions $\bar{x}_1(\bar{t})$ and $\bar{x}_2(\bar{t})$ of the corresponding TDHOE can be used to form a single complex-valued solution: $\varphi = \bar{x}_1(\bar{t}) + i \cdot \bar{x}_2(\bar{t}) = \rho \cdot e^{i\theta}$, where $\rho = \sqrt{\bar{x}_1^2 + \bar{x}_2^2}$ and $\theta = \arctan\left(\bar{x}_2/\bar{x}_1\right)$. Alternatively, if the complex-valued solution of the TDHOE is known, its real and imaginary parts represent two linearly independent real-valued solutions of the same equation. The complex-valued function $\varphi(\bar{t})$ coincides exactly with the dimensionless wavefunction $\bar{\psi}(\bar{x})$ of the corresponding quantum tunneling problem, thus establishing a link between two problems and providing an opportunity to describe phenomena in classical domain in terms of those in quantum domain and vice versa. In the case of quantum tunneling, we know that $\left|\bar{\psi}(\bar{x})\right|^2 d\bar{x}$ gives the *relative probability* of finding a quantum particle in the interval $[\bar{x}, \bar{x} + d\bar{x}]$. Let us now show that, for the corresponding TDHO, the quantity $\left|\varphi(\bar{t})\right|^2 d\bar{t}$ has analogous interpretation. Indeed, a radius-vector of a point $C = (\bar{x}_1(\bar{t}), \bar{x}_2(\bar{t}))$ moving along some trajectory in a two-dimensional plane $(\bar{x}_1, \bar{x}_2)$ always traces the same area per unit of dimensionless time according to the law:

$$\rho^2 \dot{\theta} = W = \bar{x}_1 \cdot \dot{\bar{x}}_2 - \dot{\bar{x}}_1 \cdot \bar{x}_2 = \left|A\right|^2 (1 - R) = const. \,, \tag{45}$$

where $W$ is the Wronskian (see section 2). It follows that the closer point $C$ is to the origin of coordinates (i.e., the smaller $\rho$), the faster it moves to satisfy condition (45). Let us select two segments of the trajectory around two distinct amplitudes $\rho_1$ and $\rho_2$ both lying within small phase intervals of equal magnitudes: $[\theta_1, \theta_1 + \delta\theta]$ and $[\theta_2, \theta_2 + \delta\theta]$, respectively. For small enough $\delta\theta$, the amount of time the point spends within the segment $k$ ($k$ = 1, 2) is given by $\tau_k \approx \dfrac{\delta\theta}{\dot{\theta}_k} = \dfrac{\delta\theta}{W} \cdot \rho_k^2$ . The ratio $\tau_1/\tau_2$ equals to the ratio of probabilities of finding the point $C$ within the corresponding segments. Specifically, we have

$$\frac{P\{C \in Segment\ 1\}}{P\{C \in Segment\ 2\}} = \frac{\tau_1}{\tau_2} = \frac{\rho_1^2}{\rho_2^2} \tag{46}$$

Hence $\left|\varphi(\bar{t})\right|^2 d\bar{t} = \rho^2 d\bar{t}$ yields the *relative probability* of 'observing' the point $C$ at distance $\rho$ from the origin (0, 0) within time interval $[\bar{t}, \bar{t} + d\bar{t}]$. Thus we see that there exists a remarkable correspondence between quantum tunneling problem in one dimension and behavior of two one-dimensional classical TDHOs. Specifically, the trajectories of two uncoupled classical oscillators, which initially have a 90° relative phase shift and then are perturbed parametrically and simultaneously by the same protocol, can be used to form a complex-valued function which exactly coincides with the wavefunction describing quantum tunneling through a potential barrier of the corresponding shape. The opposite is also true: given a wavefunction $\bar{\psi}(\bar{x})$ – dimensionless solution of TISE for a quantum tunneling problem, one can thought of its real and imaginary parts as trajectories of two uncoupled TDHOs parametrically perturbed at the same time using the protocol related to the potential barrier shape by (23). Figure 8 illustrates this relationship between classical and quantum domains using the results obtained earlier (Example 3).

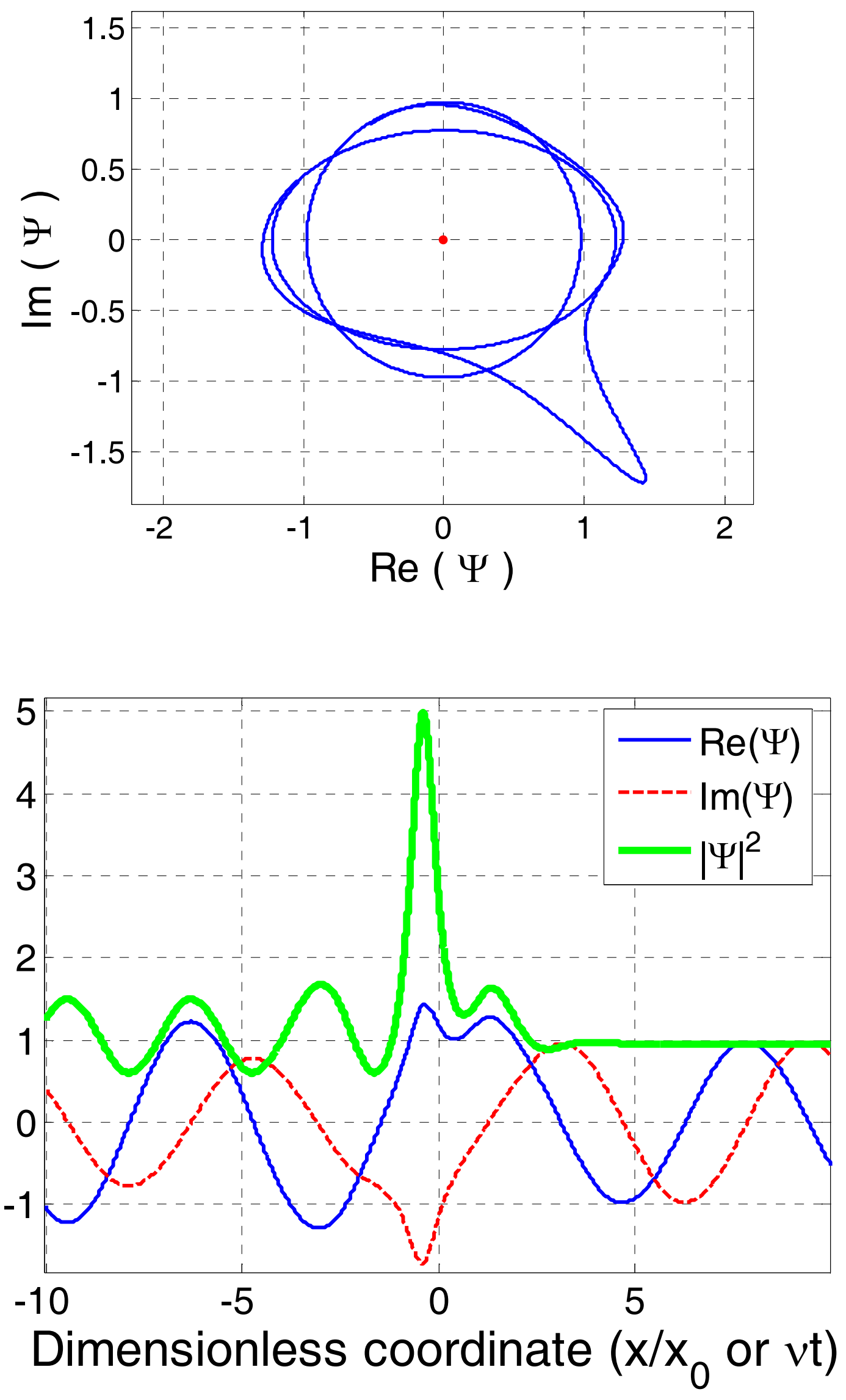

**Figure 8:** Results for Example 3 presented in a way that reveals the link between quantum and classical physics, see text for details. Top - The trajectory of a point in a plane (Re( $\psi$ ), Im( $\psi$ )). Bottom – Real and imaginary components of $\psi$ along with $|\psi|^2$ as functions of dimensionless coordinate $x/x_0$ (for quantum tunneling) or $\nu t$ (for corresponding TDHO).

## 8 Conclusions

The relationship between quantum and classical physics is still far from being completely understood. The results obtained herein shed new light on this relationship and suggest that, at least in the case of one dimension, language adopted in classical domain can be utilized to fully *describe* (but not interpret) quantum phenomena and vice versa. Such possibility has been clearly demonstrated in the case of one-dimensional quantum tunneling problem. These results are based on a novel method of finding a pair (TDHO equation, its exact solution) which can be employed for generating many previously unknown such pairs. The method is also applicable to one-dimensional propagation of electromagnetic waves through inhomogeneous media with many important applications. The conjecture has been put forward that a classical (macroscopic) harmonic oscillator disturbed by an arbitrary *parametric* perturbation of a finite duration can manifest behavior characterized by the 'range relations' analogues to the time-energy and position-momentum uncertainty relations of quantum mechanics, although with different interpretation.